\documentclass[prl,twocolumn,nofootinbib]{revtex4}
\usepackage{graphicx}
\usepackage{dcolumn}
\usepackage{amssymb}
\usepackage{bm}
\usepackage{enumerate}
\def\be{\begin{enumerate}}
\def\ee{\end{enumerate}}
\def\beq{\begin{equation}}
\def\eeq{\end{equation}}
\def\bea{\begin{eqnarray}}
\def\eea{\end{eqnarray}}

\def\3halfs{\textstyle{\frac{3}{2}}} 
\newcommand{\sla}[1]%
        {{\raise.15ex\hbox{$/$}\kern-.57em #1}}
\newcommand{\Sla}[1]%
        {{\raise.15ex\hbox{$/$}\kern-.75em #1}}

\def\ben{\begin{enumerate}}
\def\een{\end{enumerate}}
\def\bitem{\begin{itemize}}
\def\eitem{\end{itemize}}

\def\i{{\mathrm{i}}}

\begin{document}
\def\Universita{Universit\`a}
\title{\bf \Large
On horizon constraints and Hawking radiation}

\author{David Mattingly}
\affiliation{Department of Physics, University of
California-Davis} \email{mattingly@physics.ucdavis.edu}


\begin{abstract}
Questions about black holes in quantum gravity generally
presuppose the presence of a horizon.  Recently Carlip has shown
that enforcing an initial data surface to be a horizon leads to
the correct form for the Bekenstein-Hawking entropy of the black
hole.  Requiring a horizon also constitutes fixed background
geometry, which generically leads to non-conservation of the
matter stress tensor at the horizon.  In this work, I show that
the generated matter energy flux for a Schwarzschild black hole is
in agreement with the first law of black hole thermodynamics, $8
\pi G \Delta Q = \kappa \Delta A$.
\end{abstract}

\maketitle

\section{Introduction}

In the search for a theory of quantum gravity, black holes play an
important role.  Since black holes behave as thermodynamic
objects, with an entropy given by $S=A/4 \hbar
G$~\cite{Bekenstein:1973ur,Hawking:1974rv}, they presumably are
connected to the underlying microstates of quantum gravity.
Indeed, almost every modern approach to quantum gravity makes
attempts, with varying degrees of success, to show that the number
of microstates for a black hole matches that predicted by black
hole thermodynamics (see~\cite{Wald:1999vt} for a review). With a
top-down approach this is feasible - one has the fundamental
theory and can define the particular solutions that represent a
black hole in the macroscopic limit. These solutions have a
statistical mechanics determined by the fundamental theory, and
the resulting macroscopic thermodynamics can be compared with the
laws of black hole thermodynamics.

Unfortunately, we don't have a complete theory of quantum gravity.
It is therefore useful to analyze the statistical mechanics of
black holes from a bottom up approach as well, in the hopes that
information might be derived about the physics the statistical
mechanics of the black hole microstates should follow. In doing so
we immediately run into a puzzling problem. Black hole
thermodynamics is well defined in the semiclassical approximation:
we know what spacetime represents a black hole and can investigate
the behavior of quantum fields and small gravitational
fluctuations on top of the background spacetime.  In a full theory
of quantum gravity this splitting is unacceptable. There is no
background spacetime available and metric fluctuations due to the
uncertainty principle will prevent the localization of a horizon.
Now our problem becomes clear: if we cannot even identify the
states that represent a black hole, how can we possibly analyze
their statistical mechanics?

Two approaches have been developed for dealing with this issue.
The first is the ``horizon as boundary''
approach~\cite{Carlip:1996mi} which treats the black hole horizon
as a boundary $H$.  Imposing boundary conditions on the path
integral for the metric inside and outside the horizon such that
$H$ is a horizon yields the appropriate number of boundary states.
Another, more recent, alternative is to impose constraints
directly on the theory such that $H$ is a horizon.  In a recent
work, Carlip has shown that imposition of ``horizon constraints''
yields the appropriate number of microstates for black hole
entropy~\cite{Carlip:2004mn}.  In this approach $H$ is a spacelike
initial data surface constrained to be a stretched horizon.  For a
review of both these approaches see
also~\cite{Carliptobepublished} and references therein.

The notion of a stretched horizon is important for this work, so I
now take a little time to explain what it is and why it is useful.
Naively, one would want $H$ to be a true black hole event horizon.
This is problematic however as the event horizon is a global
object and requires knowledge about the entire spacetime as
opposed to a local picture for microstates.  A local definition of
horizon which still captures black hole thermodynamics is that of
an isolated horizon~\cite{Ashtekar:1998sp} (see
also~\cite{Ashtekar:2004cn, Booth:2005qc} for overviews of the
isolated horizons program). An isolated horizon is essentially
just a null hypersurface with vanishing expansion of the null
normal.\footnote{Isolated horizons also have some other
properties, one of which is that all the field equations are
satisfied on the horizon. In the approach presented here, the
classical matter field equations are deliberately not satisfied on
$H$. Therefore $H$ is not a true isolated horizon, however the
geometric characteristics are the same.} This is still not quite
good enough, as the approach of Carlip requires that $H$ be an
initial data surface and hence spacelike. If, however, one takes
$H$ to be a spacelike stretched horizon, which is defined as a
nearly null spacelike hypersurface that traces a true isolated
horizon, then one can use the Hamiltonian formulation of general
relativity.

Assuming that $H$ is a stretched horizon, Carlip looks at the
algebra of diffeomorphisms on $H$ in 2-d dilaton gravity.  The
usual generators of translations along the surface and local
Lorentz boosts in general relativity do not preserve the horizon
constraints.  If $H$ is to be a horizon, however, then the
generators \textit{must} preserve the horizon constraints, which
means that they should be modified from their usual form. The
necessary modification of the generators leads to a generator
algebra equivalent to that of a conformal field theory with
non-zero central charge and a calculation of the density of states
using the Cardy formula yields the standard result for the entropy
of a black hole.

This result is intriguing as it tells us about the density of
states from the classical symmetries of the action, rather than
some specific quantum gravity model. However, it also suffers some
limitations. The primary shortcoming is that the constraint
approach in the Hamiltonian formulation cannot handle dynamics (as
the initial data surface is the horizon) and therefore cannot
yield information about black hole evolution. In contrast, the
first law of black hole thermodynamics is specifically about
dynamics. In this paper, I show how the imposition of horizon
constraints for a dynamical Schwarzschild black hole generates a
matter energy-momentum flux at the horizon that agrees with the
first law of black hole thermodynamics.

Intuitively how does this happen?  Let us step back and consider
more generally what happens to the structure of general relativity
when we require $H$ to be a horizon.  Relativity is often labelled
as ``diffeomorphism invariant'' and, of course, it is.  However,
special relativity is also diffeomorphism invariant (as is almost
any reasonable theory of physics that we have) since it produces
the same observable results no matter what coordinate system one
uses to calculate in.  The real way in which general relativity is
unique is that there are no background geometrical objects, unlike
special relativity where the spacetime metric is fixed to be the
Minkowski metric. The fact that every field in the action is
dynamical has important consequences;the most relevant of which
for this paper is that it leads to the conservation of the matter
stress tensor via Noether's theorem.

The constraint approach violates this tenet of general relativity.
The requirement that $H$ is a horizon constitutes a fixed
background geometry and hence will generally lead to a
conservation anomaly - the matter stress tensor cannot be
conserved at the horizon. \footnote{For other examples of how
anomalies can be used in black hole physics
see~\cite{Robinson:2005pd, Christensen:1977jc}.} If the theory is
completely classical then this is a problem as the matter stress
tensor is conserved as a consequence of the matter equations of
motion. The only way out is for the matter fields to be off shell,
i.e. there must be some purely quantum mechanical effect in the
matter sector responsible for generating matter flux at the
horizon. Fortunately, in a black hole spacetime we have exactly
such an effect: Hawking radiation!

Hawking radiation from a spherically symmetric black hole reduces
the area of the horizon according to the first law of black hole
thermodynamics, $ 8 \pi G \Delta Q = \kappa \Delta A$. $\Delta Q$
is the (negative) energy flux of the current $T_{\alpha \beta}
\xi^\beta$ across the horizon $H$ (where $\xi^\beta$ is the
Killing field for which $H$ is a Killing horizon), $\kappa$ is the
surface gravity of the black hole, and $A$ is the area of the
horizon cross section. There is no guarantee, of course, that the
flux generated from requiring $H$ to be a dynamical horizon should
match this law.  However, I show that for Schwarzschild geometry
this actually is the case up to an undetermined constant, and that
the exact first law is recovered if this constant is chosen to be
unity.  I now turn to the derivation of this result.

\section{Modified Einstein equations}
Consider a local region $D$ of a manifold $M$ in a coordinate
system $x^\beta=(u,v,y,z)$. (Eventually spherical symmetry will be
assumed and $y,z$ will be the angular directions.) I can define a
hypersurface $H$ by the condition $u=0$ and a vector field
$h^\alpha$ such that $h^\alpha|_H$ generates $H$. One should think
of $h^\alpha$ as the (approximate) Killing field for which $H$ is
to be a Killing horizon. Furthermore, define a covector field
$i_\alpha$, proportional to $\partial /\partial v$, such that on
$H$ I have $i \cdot h=-1$. This construction is metric independent
since I have not yet imposed any restrictions on $g_{\alpha
\beta}$.  I remind the reader that $h^\alpha$ and $i_\alpha$ are
not to be considered dynamical fields, but rather fixed background
geometrical objects.

I now must decide how to impose the constraints that make $H$ a
horizon.  A horizon emitting Hawking radiation is not quite null,
but instead slightly timelike~\cite{Ashtekar:2005cj}.  Therefore
the Hamiltonian formulation of Carlip is not directly applicable
as $H$ cannot be an initial data surface. Instead I shall work in
the Lagrangian formulation and enforce the constraints via
Lagrange multipliers in the action.  I take $H$ to be a timelike,
instead of spacelike, stretched horizon (a timelike membrane in
the language of~\cite{Ashtekar:2004cn}) . As well, I will not
require the expansion $\Theta$ of $h^\alpha$ to identically vanish
but instead be infinitesimally small,which will give an
infinitesimal change in the cross sectional area of the horizon.
Hence the constraints are
\begin{enumerate}
    \item $\Phi_h=h^\alpha h_\alpha+L^2=0$
    \item $\Phi_\Theta = P^{\alpha \beta} \nabla_\alpha h_\beta-\Theta=0$
\end{enumerate}
where $L, \Theta$ are infinitesimal on $H$ and $P^{\alpha \beta}$
is the projection tensor $P^{\alpha \beta} = g^{\alpha \beta} -
h^\alpha h^\beta/(h \cdot h)$.

So far I have not said anything about the surface gravity of the
horizon, which also needs to be specified when analyzing the
dynamics of a black hole.  If $H$ was a true null surface then $L$
would be zero and constant along $H$, which implies that the the
directional derivative of $h^2$ would be of the usual form
\begin{equation}
\nabla_\alpha h^2 = 2 \kappa  h_\alpha
\end{equation}
where $\kappa$ is the surface gravity. Since $h^\alpha$ is not
quite null $\nabla_\alpha h^2$ is not proportional to $h^\alpha$.
In spherical symmetry $\nabla_\alpha h^2$ can be written as
\begin{equation}
\nabla_\alpha h^2 = 2 \kappa h_\alpha + \rho i_\alpha
\end{equation}
where $\rho$ is a function that depends on the evolution of the
surface. Contracting with $i^\alpha$ gives
\begin{equation} \label{eq:idoth}
i^\alpha \nabla_\alpha h^2 = -2 \kappa + \rho i^2.
\end{equation}

If I impose the constraint $i^2=L^2$ then as $L \rightarrow 0$ the
$i^2$ term drops out of (\ref{eq:idoth}).   So, finally, I fix the
surface gravity $\kappa$ by imposing two additional constraints on
$H$,
\begin{enumerate}
    \item $\Phi_i=i^\alpha i_\alpha-L^2=0$
    \item $\Phi_\kappa = i^\alpha \nabla_\alpha h^2 + 2 \kappa=0$
\end{enumerate}
which gives the right value for $\kappa$ on the constraint
surface. There is still an ambiguity in $i^\alpha$ as the above
conditions do not uniquely fix $i_\alpha$ on $D$. I use this
freedom to further choose $i^\alpha$ such that it satisfies the
geodesic equation, $i^\beta \nabla_\beta i_\alpha =0$.  Note that
this last choice for $i^\alpha$ is not a constraint as it is not a
consequence of the presence of a horizon with surface gravity
$\kappa$. Therefore it will not be implemented with a specific
constraint term in the action. Instead, given a solution for
$g_{\alpha \beta}$ and $i_\alpha$ on $H$ I simply choose
$i_\alpha$ to satisfy the geodesic equation everywhere.

The constraints are to be imposed only on $H$.  Using the Dirac
delta function, defined by
\begin{equation} \label{eq:defd}
\int d^nx du F(u,x) \delta(u)= \int d^{n}x F(0,x)
\end{equation}
the constraints can be enforced only on $H$ by adding them to the
action,
\begin{eqnarray} \label{eq:action}
S=\int \sqrt{-g}  d^4x [R + 8 \pi G \mathcal{L}^{(m)}(\Psi)
\\ \nonumber  - \delta(u) (\lambda_i \Phi_ i + \lambda_h \Phi_h + \lambda_\Theta
\Phi_\Theta + \lambda_\kappa \Phi_\kappa)]
\end{eqnarray}
where $\mathcal{L}_m(\Psi)$ is the matter Lagrangian (as a
function of matter fields $\Psi$) and the $\lambda$'s are Lagrange
multipliers.

The choice of volume element for the constraint terms requires
some explanation.  Integrating over $u$, the extra constraint
terms can be rewritten as
\begin{equation}
-\int_H d^3x \sqrt{-g}(\lambda_i \Phi_ i + \lambda_h \Phi_h +
\lambda_\Theta \Phi_\Theta + \lambda_\kappa \Phi_\kappa).
\end{equation}
At first glance, this term seems wrong as the integration over $H$
naturally involves the determinant of the induced metric
$q_{\alpha \beta}$ on $H$, rather than $\sqrt{-g}$.  However, let
us consider more carefully the effect on, for example,  the
surface gravity constraint if the volume element was chosen to be
$\sqrt{-q}$.  On $H$, the equation of motion derived from varying
$\lambda_\kappa$ is $\sqrt{-q} \Phi_\kappa=0$ which enforces
$\Phi_\kappa=0$ only if $\sqrt{-q} \neq 0$.  In the limit as $L
\rightarrow 0$, $H$ becomes null and $\sqrt{-q} \rightarrow 0$ as
well.  In this limit the constraint $\Phi_\kappa$ is completely
lost as the $\lambda_\kappa$ equation of motion is satisfied for
any value of $\Phi_\kappa$.  Therefore the natural volume element
on $H$ is \textit{not} the appropriate quantity to use for
enforcing constraints. Integrating the constraints with respect to
the four volume element $\sqrt{-g}$ does, however, enforce the
constraints as $\sqrt{-g}$ is non-vanishing everywhere (as long as
the metric is non-singular).

Varying (\ref{eq:action}) with respect to the Lagrange multipliers
enforces the constraints.  Varying (\ref{eq:action}) with respect
to $g^{\alpha \beta}$ and using the identity $h^\alpha
\nabla_\alpha \delta(u)=0$ (since $h^\alpha$ is a generator of
$H$) yields the modified Einstein equation
\begin{equation} \label{eq:EE}
G_{\alpha \beta} - C_{\alpha \beta} = 8 \pi G T^{(m)}_{\alpha
\beta}.
\end{equation}
$C_{\alpha \beta}$, the constraint ``stress tensor'', is given by
\begin{eqnarray} \label{eq:C1}
C_{\alpha \beta}= \delta(u)[\lambda_i i_\alpha i_\beta -\lambda_h
h_\alpha h_\beta + \lambda_\kappa i_{(\alpha} \nabla_{\beta)} h^2
\\ \nonumber + \frac {\lambda_\Theta} {h^2} (-h_{(\alpha} \dot{h}_{\beta)} + \frac {1}
{h^2} h_\alpha h_\beta h_\gamma \dot{h}^\gamma)] \\ \nonumber +
h_\alpha h_\beta \nabla_\gamma (\lambda_\kappa \delta(u) i^\gamma)
+ \delta(u) \frac {1} {2} \nabla_\gamma(\lambda_\Theta h^\gamma
P_{\alpha \beta})
\end{eqnarray}
where $\dot{h}^\alpha=h^\beta \nabla_\beta h^\alpha$.  Note that
in (\ref{eq:C1}) I have imposed the constraints and dropped terms
that vanish (such as those from the variation of $\sqrt{-g}$) when
the constraints are satisfied. $C_{\alpha \beta}$ can be further
simplified by noting that in the limit of $L \rightarrow 0$,
$h^\alpha$ is the generator of a null hypersurface and so
$\dot{h}^\alpha = f h^\alpha$ where $f$ is some scalar function.
In this limit the $\lambda_\Theta$ terms in the second line of
(\ref{eq:C1}) cancel, leaving the simpler expression

\begin{eqnarray} \label{eq:C2}
C_{\alpha \beta}= \delta(u)[\lambda_i i_\alpha i_\beta -\lambda_h
h_\alpha h_\beta + \lambda_\kappa i_{(\alpha} \nabla_{\beta)} h^2
\\ \nonumber +  \frac {1} {2} \nabla_\gamma(\lambda_\Theta h^\gamma P_{\alpha
\beta}) ]
 + h_\alpha h_\beta \nabla_\gamma (\lambda_\kappa
\delta(u) i^\gamma).
\end{eqnarray}
I now discuss how the presence of $C_{\alpha \beta}$ in the field
equations affects the conservation of the matter stress tensor.

\section{Generated matter flux}
Let us consider a simple black hole geometry, that of a
Schwarzschild black hole. Hence from this point on spherical
symmetry is assumed and all quantities are independent of $y,z$.
$G_{\alpha \beta}$ vanishes and so contracting (\ref{eq:EE}) with
$h^\beta$ and taking the divergence yields

\begin{equation} \label{eq:Div}
-\nabla^\alpha (C_{\alpha \beta} h^\beta) = 8 \pi G \nabla^\alpha
(T^{(m)}_{\alpha \beta} h^\beta).
\end{equation}

\begin{figure}[htb]
\scalebox{0.40} {\includegraphics{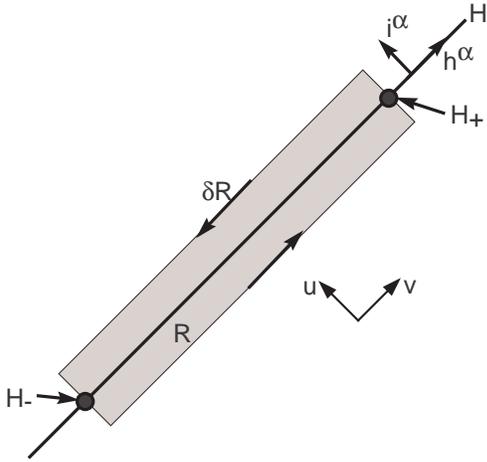}}
\caption{Horizon geometry and the surface $R$.  Each point
represents a 2-sphere in $(y,z)$.}
\label{fig:fig1}
\end{figure}

I now integrate (\ref{eq:Div}) over an infinitesimally wide region
$R$ that contains a portion of $H$ (Fig.~\ref{fig:fig1}).
Since the terms in (\ref{eq:Div}) are total divergences Stokes'
theorem gives
\begin{equation} \label{eq:intdiv}
-\int_{\partial R} \sqrt{\gamma} n^\alpha C_{\alpha \beta} h^\beta
= 8 \pi G \int_{\partial R} \sqrt{\gamma} n^\alpha T^{(m)}_{\alpha
\beta} h^\beta
\end{equation}
where $n^\alpha$ is the normal to the boundary $\partial R$ of $R$
and $\sqrt{\gamma}$ is determinant of the induced metric
$\gamma_{\alpha \beta}$ on $\partial R$. I choose $\partial R$
such that $i^\alpha$ generates $\partial R$ across $H$. Since
$i_\alpha$ is proportional to $\partial / \partial v$ and I have
the constraints $i^2, h^2 \approx 0, i \cdot h = -1$, $i^\alpha$
generates translations in the $u$ direction. I will therefore be
integrating over $u$ and so can easily evaluate the integral over
$\delta(u)$. The right hand side of (\ref{eq:intdiv}) is simply $8
\pi G$ times the total matter flux $\Delta Q$ emanating from the
horizon through the surface $\partial R$. Without the constraints
this would be zero, corresponding to no anomalous flux for the
approximate Killing field $h^\alpha$.

I now evaluate the left hand side of (\ref{eq:intdiv}). $C_{\alpha
\beta} h^\beta$ is
\begin{eqnarray} \label{eq:Cdoth}
C_{\alpha \beta} h^\beta = \delta(u) [ -\lambda_h h_\alpha h^2
-\lambda_i i_\alpha \\ \nonumber +\frac {1} {2} \lambda_\kappa
(i_\alpha h^\beta \nabla_\beta h^2 -\nabla_\alpha h^2) - \frac {1}
{2} \lambda_\Theta P_{\alpha \beta} \dot{h}^\beta] \\
\nonumber + h_\alpha h^2 \nabla_\gamma (\lambda_\kappa \delta(u) i
^\gamma)
\end{eqnarray}
where I have applied the Liebniz rule to put the $\lambda_\Theta$
term in the above form.  In the limit as $h^\alpha$ is null, the
$\lambda_\Theta$ term vanishes. The presence of the delta
functions implies that there is only a contribution to the
integral when $\partial R$ crosses $H$ (the two dimensional
surfaces $H_-,H_+$ in Figure \ref{fig:fig1}).  Therefore I can
safely use the identity $n \cdot i=0$. Contracting with $n^\alpha$
and applying $n \cdot i =0$, I have
\begin{eqnarray} \label{eq:ndotCdoth}
n^\alpha C_{\alpha \beta} h^\beta = - \delta(u) \lambda_h h^2 (n
\cdot h) \\ \nonumber - \frac{1} {2} \delta(u) \lambda_\kappa
n^\alpha \nabla_\alpha h^2 + (n \cdot h) h^2 \nabla_\gamma
(\delta(u) \lambda_\kappa i^\gamma).
\end{eqnarray}
I now rewrite the last term in (\ref{eq:ndotCdoth}) to move the
delta function outside the derivative. Using the identity
$\delta_\alpha^\beta = \gamma_\alpha^\beta - n_\alpha n^\beta$ the
last term can be rewritten as
\begin{eqnarray} \label{eq:kappaterm}
(n \cdot h) h^2 \nabla_\gamma (\delta(u) \lambda_\kappa i^\gamma)
\\ \nonumber = (n \cdot h) h^2 [D_\gamma (\delta (u) \lambda_\kappa
i^\gamma) - \delta(u) \lambda_\kappa n^\alpha n^\beta \nabla_\beta
i_\alpha]
\end{eqnarray}
where $D_\alpha = \gamma_\alpha^\beta \nabla_\beta$.  The last
term in (\ref{eq:kappaterm}) vanishes in the limit $L \rightarrow
0$ as $n^\alpha \approx \pm i^\alpha / \sqrt{i^2}$ and (as chosen
previously) $i^\alpha$ satisfies the geodesic equation. The
plus/minus is present as $n^\alpha$ changes orientation from $H_+$
(where $n^\alpha$ is approximately parallel to $i^\alpha$) to
$H_-$ (where the two vectors are approximately anti-parallel). The
$D_\gamma$ term in (\ref{eq:kappaterm}) can be integrated over
$\partial R$ by parts and becomes
\begin{equation}
- \delta(u) \lambda_\kappa i^\gamma D_\gamma ((n \cdot h) h^2).
\end{equation}
There is no boundary term due to the presence of the delta
function. The entire integral in (\ref{eq:intdiv}) is then
proportional to $\delta(u)$. Integrating over the delta function
(\ref{eq:intdiv}) becomes
\begin{eqnarray}
8 \pi G \Delta Q= -\int_{H_+, H_-} \sqrt{\gamma}[ \lambda_h h^2 (n
\cdot h)
 - \frac{1} {2}  \lambda_\kappa n^\alpha
\nabla_\alpha h^2 \\ \nonumber - \lambda_\kappa i^\gamma D_\gamma
((n \cdot h) h^2)]
\end{eqnarray}
where the notation $\int_{H_+, H_-}$ signifies that the result is
the sum of the integrals over $H_+$ and $H_-$.  Again replacing
$n^\alpha$ by $\pm \i^\alpha/\sqrt{i^2}$ one sees that the
$\lambda_h$ term vanishes as $h^2 \rightarrow 0$.  Applying the
constraint $\Phi_\kappa$, $h^2 \rightarrow 0$, and (again) the
fact that $i^\alpha$ satisfies the geodesic equation the last two
terms combine and I have
\begin{equation}
8 \pi G \Delta Q=  \int_{H_+} \sqrt{\gamma}  \lambda_\kappa \frac
{\kappa} {\sqrt{i^2}} - \int_{H_-} \sqrt{\gamma} \lambda_\kappa
\frac {\kappa} {\sqrt{i^2}}
\end{equation}
Now note that in this coordinate system $\sqrt{\gamma}$ is equal
to $\sqrt{i^2} \sqrt{\gamma_2}$, where $\gamma_2$ is the induced
metric on $H_+,H_-$.  The factors of $\sqrt{i^2}$ cancel and I am
finally left with
\begin{equation}
8 \pi G \Delta Q= \kappa (\int_{H_+} \sqrt{\gamma_2}
\lambda_\kappa -  \int_{H_-} \sqrt{\gamma_2} \lambda_\kappa).
\end{equation}
If $\lambda_\kappa$ varies between $H_+$ and $H_-$ then there is a
non-zero $\Delta Q$ even if the geometry does not change. However,
this does not correspond to the case of interest, where the flux
is due to the time dependence of the geometry and not the time
dependence of the Lagrange multipliers.  Therefore
$\lambda_\kappa$ should be chosen to be a constant.  If I choose
$\lambda_\kappa=1$ then I recover that the generated matter flux
satisfies the first law of black hole thermodynamics,
\begin{equation}
8 \pi G \Delta Q= \kappa (\int_{H_+} \sqrt{\gamma_2} - \int_{H_-}
\sqrt{\gamma_2}) = \kappa \Delta A
\end{equation}
which is the promised result.  In summary, I have shown that
requiring the presence of a horizon with surface gravity $\kappa$
by imposing constraints in the gravitational action also generates
a (necessarily quantum mechanical) matter flux which, at least for
Schwarzschild geometry, agrees with the first law of black hole
thermodynamics.

\section{Discussion}
There are some open issues with the above result.  The most
pressing is obviously the value of $\lambda_\kappa$.  While the
choice $\lambda_\kappa=1$ is not an exceptional or fine tuned
value, there is no \textit{a priori} reason why this choice is the
correct one.  The difficulty in determining $\lambda_\kappa$ stems
from the fact that I am using the variation of the horizon to
determine the generated matter flux without reference to any
matter equations of motion. If the (quantum mechanical) matter
response was known then it is possible that the value of
$\lambda_\kappa$ could be fixed by applying the constraint.
However, without the matter field equations this method of
determining $\lambda_\kappa$ is not applicable.

Another open issue is the connection with Carlip's method for
calculating black hole entropy.  One would hope that a connection
can be made since the underlying approach, applying constraints to
force a horizon, is the same. If a connection can be made and in
the framework presented here the entropy can be shown to be $A/4$,
then the first law is a true thermodynamic equation since in this
derivation it relates the outgoing energy flux to the number of
microstates of the full quantum theory.  There are a few obstacles
that still need to be overcome in order to relate the two
approaches. Since the horizon surface is a timelike hypersurface
here (as it must be to emit Hawking radiation), it is somewhat
unclear how to proceed with the Hamiltonian framework.
Furthermore, the constraints on the surface gravity do not appear
in the formalism of Carlip, so there is no guarantee that the
algebra of constraints is identical. If these technical problems
can be overcome, then the constraint program might yield not only
the number of microstates of a black hole, but also more directly
tie Hawking radiation and the laws of black hole thermodynamics to
the microscopic theory of quantum gravity.

\section*{Acknowledgements}
I heartily thank Steve Carlip and Damien Martin for many helpful
discussions and for reading drafts of this manuscript.  This work
was funded under DOE grant DE-FG02-91ER40674.

\end{document}